\documentclass[journal]{IEEEtran}

\usepackage{cite}
\ifCLASSINFOpdf
  \usepackage[pdftex]{graphicx}
\else
  \usepackage[dvips]{graphicx}
\fi

\usepackage{amsmath}
\usepackage{array}

\ifCLASSOPTIONcompsoc
  \usepackage[caption=false,font=normalsize,labelfont=sf,textfont=sf]{subfig}
\else
  \usepackage[caption=false,font=footnotesize]{subfig}
\fi
\usepackage{subfig}
\usepackage{caption}
\usepackage[hyphens]{url}
\begin{document}

\title{To be a pro-vax or not, the COVID-19 vaccine conundrum on Twitter}
%
%
%

\author{Zainab~Zaidi,~\IEEEmembership{Senior Member,~IEEE,}
        Mengbin~Ye,~\IEEEmembership{Member,~IEEE,}
        Shanika~Karunasekera,~\IEEEmembership{Member,~IEEE}
        and~Yoshihisa~Kashima
        \vspace{-10mm}
\thanks{Email: zzaidi@ieee.org}
\thanks{Manuscript received ; revised .}}

\markboth{Journal of \LaTeX\ Class Files,~Vol.~, No.~, Mon~year}%
{Zaidi \MakeLowercase{\textit{et al.}}: Bare Demo of IEEEtran.cls for IEEE Journals}
%

\maketitle

\begin{abstract}
The most surprising observation reported by the study in \cite{Zaidi2023}, involving stance detection of COVID-19 vaccine related tweets during the first year of pandemic, is the presence of a significant number of users ($\sim$2 million) who posted tweets with both anti-vax and pro-vax stances. This is a sizable cohort even when the stance detection noise is considered. In this paper, we tried to get deeper understanding of this {\it dual-stance} group. Out of this group, 60\% of users have more pro-vax tweets than anti-vax tweets and  17\% have same number of tweets in both classes. The rest have more anti-vax tweets, and they were highly active in expressing concerns about mandate and safety of a fast-tracked vaccine, while also tweeted some updates about vaccine development. The leaning pro-vax group have opposite composition: more vaccine updates and some posts about concerns. It is important to note that vaccine concerns were not always genuine and had a large dose of misinformation. 43\% of the balanced group have only tweeted one tweet of each type during our study period and are the less active participants in the vaccine discourse. Our temporal study also shows that the change-of-stance behaviour became really significant once the trial results of COVID-19 vaccine were announced to the public, and it appears as the change of stance towards pro-vax is a reaction to people changing their opinion towards anti-vax. Our study finished at Mar 23, 2021 when the conundrum was still going strong. The dilemma might be a reflection of the uncertain and stressful times, but it also highlights the importance of building public trust to combat prevalent misinformation.
\end{abstract}


%
\IEEEpeerreviewmaketitle

\vspace{-5mm}

\section{Introduction}
%
%
%
%
\begin{quote}{\it
\IEEEPARstart{``I}{}work in pharma and I'm not getting it either. I know how to read journal articles and clinical trial data. The risk versus potential reward isn't worth it, especially with the unknowns of an RNA vaccine.  But I'm still called a conspiracy theorist." }
\end{quote}
\begin{quote}{\it``As the virus gets worse, people get more nonchalant about it. It makes no sense.  We have vaccines. There’s a light at the end of the tunnel. Be safe for another few months and we’ll be out of this. We can all be heroes just by being careful. Saving lives has never been easier!"}
\end{quote}
The first message arguably expresses an anti-vaccination (anti-vax) stance, whereas the second message conveys a pro-vaccination (pro-vax) stance. The above two messages were actually tweets sent by the same user on Twitter (now X), and our work~\cite{Zaidi2023} indicates such users, who tweeted both pro- and anti-vax messages, are far from isolated occurrences. 

In the past two decades, there has been significant attention focused on the study of public discourse on vaccinations, including the nature of anti-vax and pro-vax content. Interest has increased especially since the COVID-19 pandemic began in 2020. Much of the existing literature has focused on the contrast between anti-vax and pro-vax discourse, and polarisation among users into distinct anti-vax and pro-vax groups \cite{Rathje2022,DeNicola2023,Monsted2022}. However, our recent study has suggested that the discourse is far less clear-cut, at least on Twitter~\cite{Zaidi2023}. In fact, the majority of tweets are posted by users showing ambivalent attitude towards COVID-19 vaccines as shown by the above example. 

Our study in Ref.~\cite{Zaidi2023} was focused on the comprehensive synoptic analysis of anti-vax and pro-vax discourse through topic modelling. The study investigated a dataset of $75$~million COVID-19 vaccine related English tweets from $11$~million users, gathered from March 2020--March 2021. We used OpenAI's GPT based stance detection tool to classify the tweets into anti-vax, pro-vax, or neutral stances. While the primary focus of \cite{Zaidi2023} was on the topics discussed, it was also found that a majority, $62\%$, of Twitter users posting anti-vax content also tweeted in favour of vaccination. These {\it dual-stance} users were sizable and prolific. Dual-stance users constituted $22\%$ of the unique users posting anti- or pro-vax tweets in our dataset, but contributed a majority of tweets in the study, 66\% and 85\% of the total pro-vax and anti-vax tweets, to be precise. This observation is in contrast to the popular perception that anti-vax and pro-vax groups over social media are highly polarized, and that such groups interact mostly with other users with similar views \cite{Rathje2022,Jiang2021,Monsted2022}. Moreover, our study \cite{Zaidi2023} affirms that many users posting anti-vax tweets were not only exposed to, but were actively engaged with, pro-vax content. Since \cite{Zaidi2023} only provided preliminary insights into dual-stance users, there is ample opportunity to further study this cohort of users, as we aim to do in this paper. Developing an understanding of this cohort may provide important insight towards policymakers designing effective campaigns for improving vaccine favorability, including for future pandemics.  

Consequently, we have formulated the following research questions, which we investigate in this paper, to develop a deeper understanding of the puzzling phenomenon of dual-stance tweeting:
\begin{itemize}
    \item [RQ1] Are dual-stance users the result of noise in the stance detection algorithm? 
    \item [RQ2] What topics do dual-stance users talk about? Do their discussion topics provide some clues about why they tweet both pro- and anti-stances? 
    \item [RQ3] Do the temporal dynamics of dual-stance users' stance change -- whether they changed to pro-stance or anti-stance -- provide some insight? Because dual-stance users must have changed their stance at least once, if not more often, change-of-stance information may provide some insight about why they did so. In particular, we explore the following questions: 1- does their change of stance correlate with COVID-19 related events, and 2- are there causal connections between changes to pro- and anti-vax stance? 
\end{itemize}

To the best of our knowledge, no existing papers have discussed the dual-stance users explicitly. However, a few works have hinted at their presence  \cite{Gori2021,Crupi2022,Giovanni2022,Poddar2022,Weinzierl2022}.  We will discuss the related work in Section~\ref{sec:related-work} in detail. Our previous paper, Ref.~\cite{Zaidi2023}, reported the presence of dual-stance users and showed via basic analysis that, even when accounting for noise in the stance-detection algorithm, there were simply too many dual-stance users who sent tweets of putatively opposing stances for this to have occurred by chance. In this paper, we provide a more detailed treatment of the issue, and show that it is possible that a user with only a few anti-vax or pro-vax tweets is falsely classified as dual-stance due to detection noise. However, there are also many users in the dataset with a significantly large number of anti-vax and pro-vax tweets, and for these users, dual-stance tweeting cannot be attributed to just detection noise, but is in fact a robust phenomenon. Details are given in Sections~\ref{sec:prep} and \ref{sec:class}.

Our analysis also shows that {\it anti-leaning} dual-stance users -- those dual-stance users who sent more anti-vax than pro-vax tweets -- were very active in expressing concerns about the vaccine mandate and safety of the vaccine, but were also posting some vaccine updates. In contrast, {\it pro-leaning} dual-stance users -- those who sent more pro- than anti-vax tweets -- have a mirror-image profile. They tweeted more about vaccine updates, but also commented on the concerns. 
 
Based on these findings, we conclude that many dual-stance users are genuine, and it appears that their stance changes reflected oscillations between pro- and anti-vax ideas under the extreme uncertainty of the COVID-19 pandemic. Although the oscillations continued as new vaccine-related events occurred throughout the pandemic, changes to a pro-vax stance became more prevalent towards the end of the study period, and by early 2023, more dual-stance users appear to have become more pro-vax than anti-vax (i.e. pro-leaning).

Moreover, our analysis suggests that stance changes occurred prominently when COVID-19 vaccine trial results were made public. The time series of changes into pro-stance and anti-stance show a high correlation with announcements of trial results. Preliminary causal analysis indicates that changes to anti-vax stance in a population trigger changes to pro-vax stance.


\section{Related Work} \label{sec:related-work}

There has been a substantial amount of work studying online public behaviour and discourse towards COVID-19 vaccines since the beginning of the pandemic. These include studies of polarisation and formation of communities and echo chambers around anti-vax and pro-vax stances \cite{Rathje2022,DeNicola2023,Monsted2022}. These studies reported that the online vaccine debate over social media was highly polarised and users only interacted with other users having similar views. Some studies focused on the correlation between political polarisation and vaccine stance \cite{Ebeling2022,Jiang2021}, and reported that users belonging to conservative ideological camps show more anti-vax attitudes. 


To our knowledge, no work has explicitly discussed dual-stance users or highlighted this phenomenon, though some publications have briefly hinted at their presence. For example, Ref.~\cite{Gori2021} manually classified 7004 Italian tweets from the period Oct 2020-Jan 2021, and assigned polarity scores to each tweet from +1 (pro-vax) to  -1 (anti-vax). Averaging across the users, they showed that the most retweeted users were only moderately polarized in the anti-vax vs pro-vax debate. Another study of Italian tweets classified users as `vaccine supporters', `vaccine hesitant', and `others' who showed no clear stance~\cite{Crupi2022}. These `others' were often communicating with the users of opposing stances, and appeared to be bridges between the polarised communities~\cite{Crupi2022}. While reporting two distinct anti-vax and pro-vax communities, Ref.~\cite{DeNicola2023} also reported the presence of {\it central} users as key players in the discourse. Ref.~\cite{Giovanni2022} categorized the hashtags of Italian, French, and German tweets from Nov 2020 to Nov 2021 into `anti-vax' and `pro-vax', and found a few tweets with both types of hashtags and termed them as `unclear'. Ref.~\cite{Poddar2022} used a BERT-based stance detection tool, trained with a combination of 1,700 labelled tweets and publicly available labels, and detected the stances of English tweets
from 4 million distinct users over the Jan 2018–Mar 2021 period. They classified users as `anti-vaxxer' or `pro-vaxxer' if {\it 70\%} of their tweets were anti-vax or pro-vax, respectively. Ref.~\cite{Weinzierl2022} used linguistic features to detect stances towards specific issues related to vaccine hesitancy. 
22\% of the users were classified into the profile `undecided', which had almost balanced stances (accept or reject) and 8\% were `concerned' users, who were supportive of vaccines but had minor concerns~\cite{Weinzierl2022}. Ref.~\cite{Kleitman2023} classified participants in a survey based study into three classes, vaccine resistant, acceptant, and moderates; the latter class showed some willingness to take vaccines as well as hesitancy. 

\section{Preliminaries}\label{sec:prep}

This paper builds upon the study presented in \cite{Zaidi2023}, which used a publicly available tweet dataset collected and maintained by R. Lamsal~\cite{Lamsal2021}. The tweets from  Mar 20, 2020 to Mar 23, 2021 were further filtered with vaccine-related keywords, and the stance of each tweet was predicted as `favour', `against', or `none' with respect to the topic `vaccine hesitancy', using the stance detection tool created from OpenAI's GPT transformer model~\cite{Rao2019}. In order to make this yearlong study possible with reasonable confidence, \cite{Zaidi2023} selected approximately $46$K+ tweets to label, sampling $100+$ tweets each day. As the public conversation on Twitter was changing throughout the year, it was critical to have a relevant labelled set of tweets to fine-tune the GPT model, which could then be used for detecting the stances for the tweets of that specific period. This fine-tuning was sufficient to give reasonable composite F-scores of 0.6 or above for 20 test sets, picked and labelled from the yearlong dataset. The composite F-scores for the 20 test sets were in the range of 0.67-0.87, precision for anti-vax stance classification was between 0.52-0.92 and that of the pro-vax class was in the range of 0.68-0.95. For details, see~\cite{Zaidi2023}. 

\subsection{Overall Statistics}\label{sec:stat}

As presented in \cite{Zaidi2023}, the stance detection tool predicted $37,047,378$ pro-vax, $10,567,955$ anti-vax, and $28,322,526$ neutral or irrelevant tweets in this dataset. Moreover, of the total of $8,637,015$ unique user IDs, $5,571,946$ ($64.5\%$) tweeted only pro-vax messages, and $1,171,837$ ($13.6\%$) tweeted only anti-vax messages, while $1,893,232$ ($21.9\%$) \emph{dual-stance} users sent out both pro- and anti-vax tweets. The dual-stance users contributed $85\%$ of the anti-vax tweets for the yearlong study period, and accounted for $62\%$  of the users who sent out anti-vax tweets (1,893,232 out of 3,053,626). On the other hand, they contributed more than half of the pro-vax tweets ($66\%$) during the study period despite only being $25\%$ of users who sent out a pro-vax tweet (1,893,232 out of 7,465,178 total pro-vax users) \cite{Zaidi2023}. 

Almost $60\%$ of the dual-stance users sent out more pro-vax than anti-vax tweets, whereas $17\%$ of the dual-stance users contributed more anti-vax than pro-vax tweets, and close to $23\%$ of the dual-stance users sent out similar numbers of anti- and pro-vax tweets during the study period. Among the dual-stance users with more pro-vax tweets, the percentage of pro-vax tweets was generally quite high; $50\%$ of these users had $80\%$ or more of their tweets as pro-vax tweets. We also compared the possibility of the dual-stance users being bots with that of users with only anti-vax and pro-vax tweets. That analysis is presented in Supplementary Material. 


\subsection{Role of Stance Detection Noise} \label{sec:st_noise}

The procedure and the results reported in \cite{Zaidi2023} give us some confidence in the observation that there exists a sizable sample of dual-stance users, and that they are not a small minority of random anomalies. 

Nevertheless, there are obviously some sources of measurement error in stance detection tools, which can cast some doubts about how prominent dual-stance users truly are. It is well known that stance detection methods, such as the one considered here, have difficulties in classifying satire and convoluted expressions. More generally, because some expressions are ambiguous, the stance classification can never be perfectly valid. For these reasons, we provide a more principled consideration of the issue of noise in stance classification, and in doing so, address Research Question~1. 

Let $p_i$ be the probability that a user $i$, classified as a dual-stance user by the stance detection algorithm, is in fact dual-stance. Noting that each tweet is classified independently into any of the three stances, and assuming that the user sent $n_a$ anti-vax and $n_p$ pro-vax tweets, $p_i$ can be estimated as below:
\begin{equation}
    p_i = 1 - (1-\alpha_a)^{n_a} - (1-\alpha_p)^{n_p} +  (1-\alpha_a)^{n_a}(1-\alpha_p)^{n_p},
\end{equation}
where $\alpha_a$ ($\alpha_p$) is the precision of the stance detection tool for the anti-vax (pro-vax) class. For our dataset $\alpha_a$ is in the range of $[0.52,~0.92]$ and $\alpha_p$ is in between $[0.68,~0.95]$ calculated using test datasets, as also described above. The effective size of the dual-stance cohort $N_{e}$ can be calculated as
\begin{equation}
    N_{e} = \sum_{i=1}^N 1.p_i,
\end{equation}
where $N$ is the number of detected dual-stance users. For the precision of stance detection method given above, $N_{e}$ is from $1,212,327$ to $1,791,219$. Even accounting for the stance detection noise, we can safely conclude that there is large number of users who have tweeted both anti- and pro-vax tweets.

\section{Classification of Dual-Stance Users}\label{sec:class}

We now explore these dual-stance users in more depth. To begin, we focus how many dual-stance users lean towards sending more pro- or anti-vax tweets, and call them {\it pro-leaning} or {\it anti-leaning} respectively. As described above, $60\%$ of the dual users have more pro-vax tweets than anti-vax tweets. This shows a tendency towards favouring COVID-19 vaccines rather than rejecting them. However, considering the imperfections in the stance detection process, it would be erroneous to classify users into pro-leaning, anti-leaning, or balanced classes based only on their number of pro-vax and anti-vax tweets. Here, {\it balanced} class refers to the users with almost similar contributions from both sides of the debate. Instead, we estimate the probability of an individual user to be in either of the above-mentioned classes. A user is subsequently placed in a class with a higher likelihood. More precisely, we defined a threshold-based assignment rule when probabilities are too close to each other.  

For tractable calculations, we assumed independent tweet classification into stances as in Section~\ref{sec:st_noise}, and also that each stance classification can result in either {\it true} or {\it false} classification, which allows us to use Binomial distributions for probability computations. The probability that a user is {\it pro-leaning} (i.e., favouring pro-vax more than anti-vax stance) is the likelihood that there are more pro-vax tweets from the user than anti-vax in the presence of stance detection errors:
\begin{IEEEeqnarray}{rCl}
    \Pr(\text{pro}) &=& \sum_{i = 1}^{min(n_a,n_p-1)}\binom{n_a}{i}\alpha_a^{i}(1-\alpha_a)^{n_a - i} \times \nonumber\\
    &&\left[\sum_{j = i+1}^{min(n_a+1,n_p)}\binom{n_p}{j}\alpha_p^{j}(1-\alpha_p)^{n_p - j} \right],
\end{IEEEeqnarray}
where $n_a$ ($n_p$) are the total number of anti-vax (pro-vax) tweets posted by the user. Similarly, the probability that a user is {\it anti-leaning} is,
\begin{IEEEeqnarray}{rCl}
    \Pr(\text{anti}) &=& \sum_{i = 1}^{min(n_p,n_a-1)}\binom{n_p}{i}\alpha_p^{i}(1-\alpha_p)^{n_p - i} \times \nonumber\\
    &&\left[\sum_{j = i+1}^{min(n_p+1,n_a)}\binom{n_a}{j}\alpha_a^{j}(1-\alpha_a)^{n_a - j} \right].
\end{IEEEeqnarray}

Also, the probability of a user being {\it balanced} is,
\begin{IEEEeqnarray}{rCl}
    \Pr(\text{bal}) &=& \sum_{i = 1}^{min(n_a,n_p)}\binom{n_a}{i}\alpha_a^{i}(1-\alpha_a)^{n_a - i} \times \nonumber\\
    &&\binom{n_p}{i}\alpha_p^{i}(1-\alpha_p)^{n_p - i}.
\end{IEEEeqnarray}

We classify each user as pro-leaning, anti-leaning, or balanced as below:
\begin{eqnarray}\label{eq:class-rule}
    \text{User is} \left \{ \begin{array}{ll}
    \text{pro-leaning}, & \text{if} \Pr(\text{pro}) > \Pr(\text{anti}) +\epsilon  \\
     & \& \Pr(\text{pro}) > \Pr(\text{bal}) +\epsilon\\
    \text{anti-leaning}, & \text{if} \Pr(\text{anti}) > \Pr(\text{pro}) +\epsilon  \\
     & \& \Pr(\text{anti}) > \Pr(\text{bal}) +\epsilon\\ 
    \text{balanced}, & \text{otherwise.}
    \end{array}\right .
\end{eqnarray}

Here, $\epsilon$ is a small tolerance value used to classify users with very close $\Pr(\text{pro})$ and $\Pr(\text{anti})$ to the balanced group. In our work, we select $\epsilon = 0.05$ (details are given in Supplementary Material). 
Figure \ref{fig:class} plots the number of anti-vax and pro-vax tweets a user sends out, for each class. A higher value of $\epsilon$ will result in a wider purple band in the middle, i.e., more users are classified as balanced. With $\epsilon = 0.05$, $50\%$ of the dual-stance users are classified as pro-leaning, $8\%$ as anti-leaning and $42\%$ as balanced. These percentages will change to $44\%$, $8\%$, and $48\%$ respectively, if instead we set $\epsilon = 0.1$. Also note that there could be many users with the same number of anti-vax and pro-vax tweets, and each point in Fig.\ \ref{fig:class} may be associated with multiple users. The balanced cohort looks smaller in Fig.\ \ref{fig:class}, although it is significantly larger in terms of number of users. 

It is important to note in Fig.~\ref{fig:class} that the balanced band is wider near the origin, or when users have a smaller number of anti- and pro-vax tweets and the impact of stance detection errors is relatively greater. For example, a user with 60 pro-vax and 30 anti-vax tweets is more likely to be a pro-leaning user than another with 2 pro-vax and 1 anti-vax tweets, though both have the same $n_p/(n_a+n_p)$ ratio. Similarly, a user with 10 anti-vax and 8 pro-vax tweets is more likely to be pro-leaning, due to higher precision of pro-vax stance prediction, than a user with 3 anti-vax and 1 pro-vax tweets, though the difference between the numbers of anti-vax and pro-vax tweets is the same in both cases. We believe that this classification method is more suitable when there is stance detection noise, rather than a simple ratio-based classification. 

Figure \ref{fig:class} also depicts some facts described in Section \ref{sec:stat}. For example, a sizable majority of tweets are pro-vax, dual-stance users tended to send more pro-vax tweets, and the percentage of pro-vax tweets was generally quite high.  

\begin{figure}
\begin{center}
\includegraphics[width= 0.46\textwidth,height=6.5cm, clip = true]{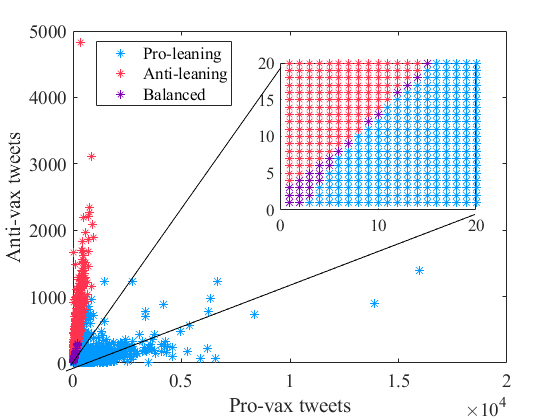}\vspace{-2mm}
\caption{Pro-leaning, anti-leaning, and balanced users with respective anti-vax and pro-vax tweets.}
\label{fig:class}
\end{center}
\end{figure}

\section{Content Analysis}\label{sec:Content-analysis} 

\subsection{Discussion Topics}

\begin{figure*}[tbh!] 
    \subfloat[]{
        \includegraphics[trim={0 0 0 0},width= 0.5\textwidth,height=7cm]{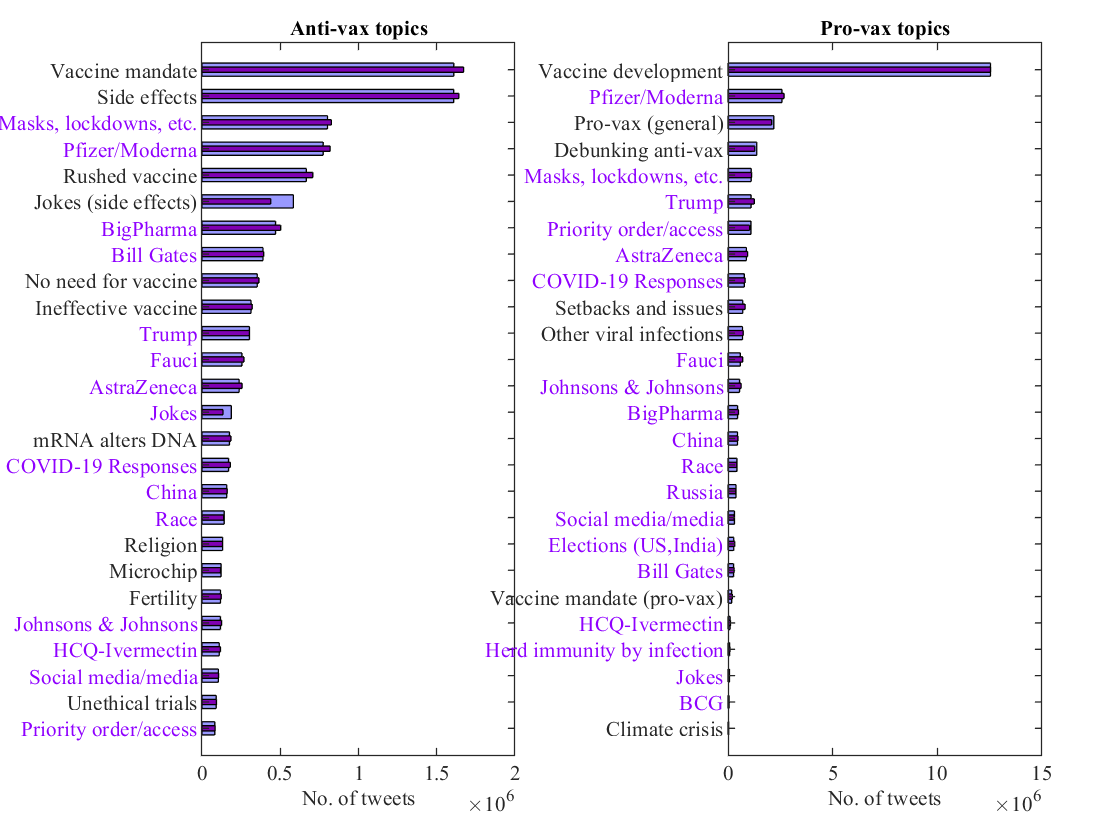}\label{topics_all}
        }
        \hfill
        \subfloat[]{        
        \includegraphics[trim={0 0 0 0},width= 0.5\textwidth,height=7cm]{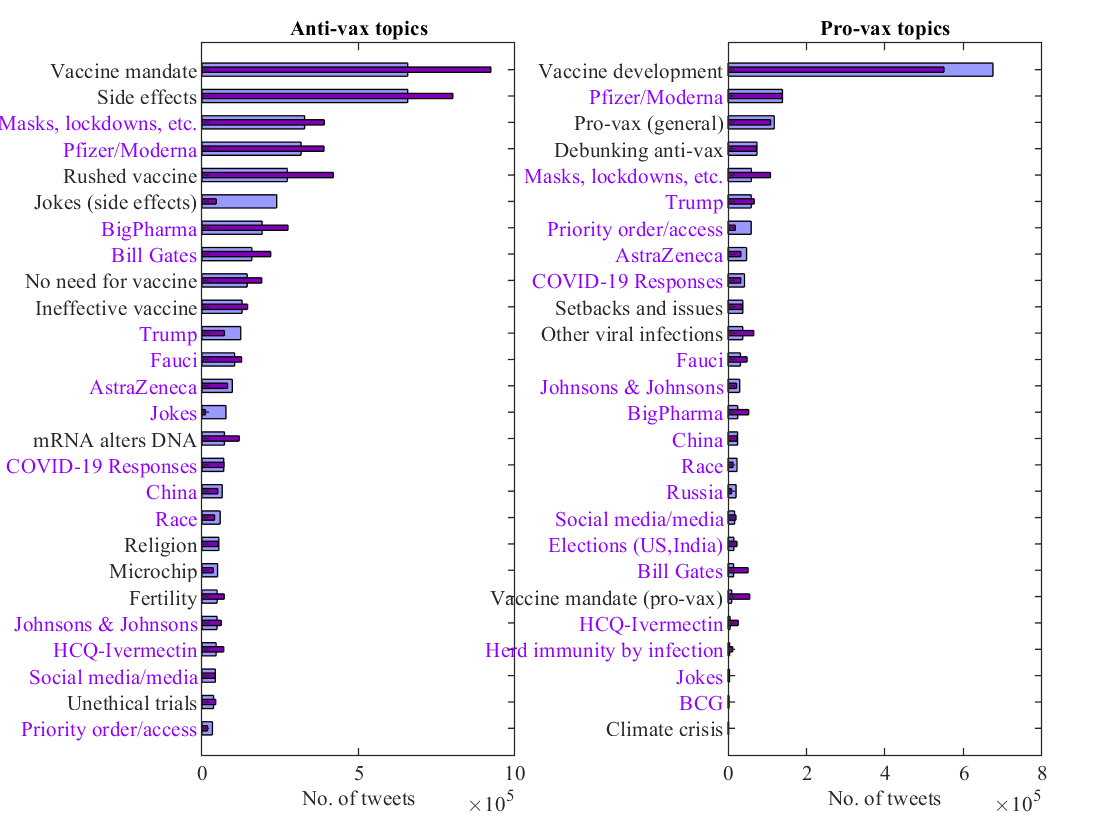}
        \label{topics_Anti}}
        \\
        \subfloat[]{
        \includegraphics[trim={0 0 0 0},width= 0.5\textwidth,height=7cm]{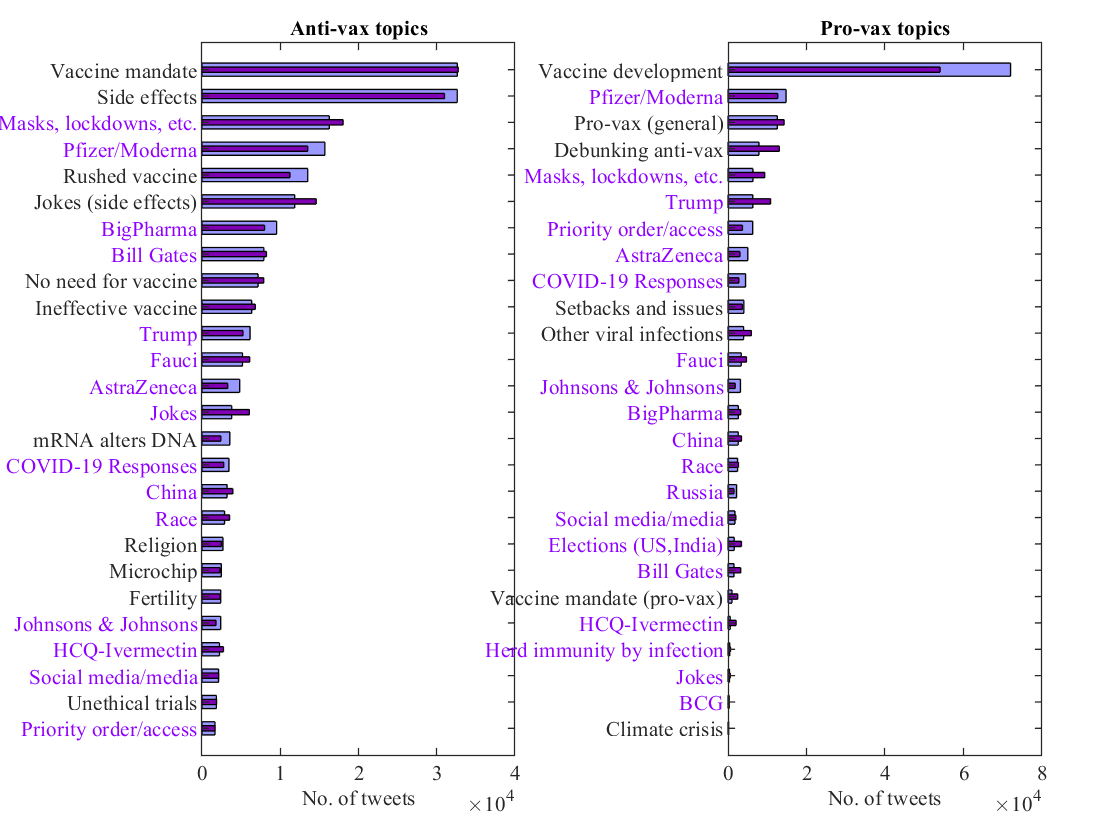}
        \label{topics_bal}}
        \hfill
        \subfloat[]{
        \includegraphics[trim={0 0 0 0},width= 0.5\textwidth,height=7cm]{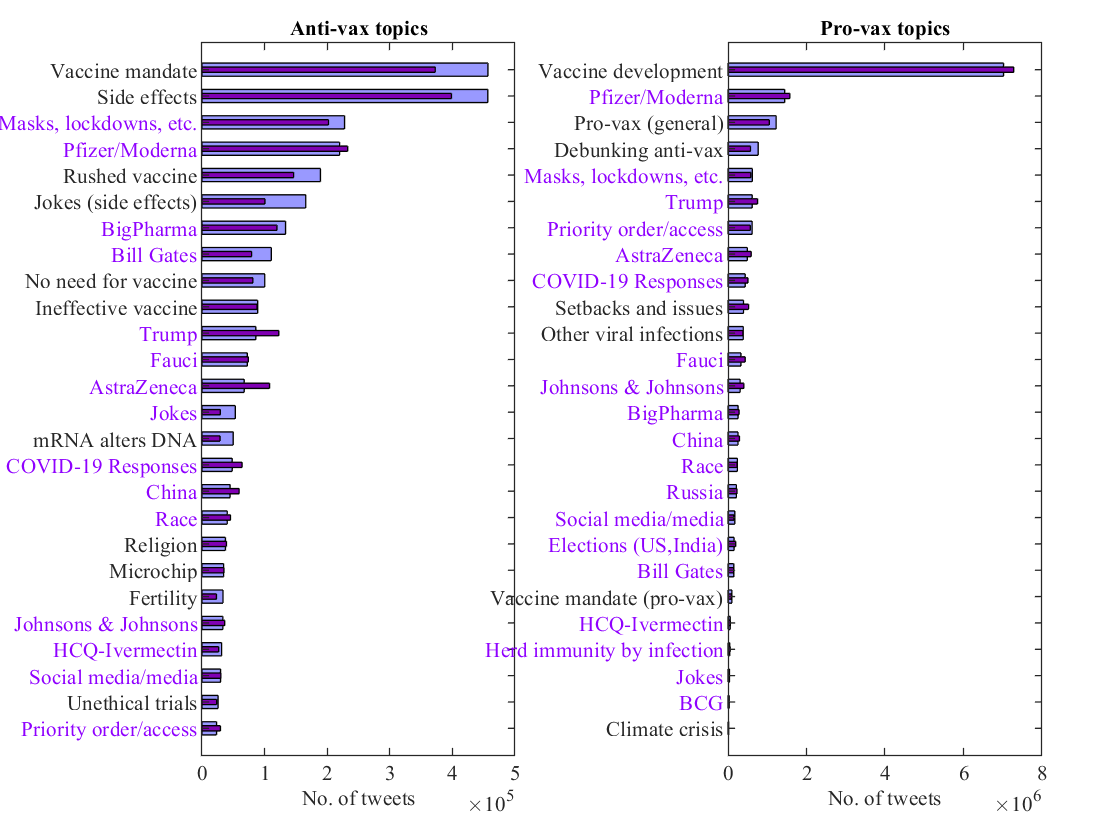}
        \label{topics_Pro}}
        \caption{Top 25 anti-vax and pro-vax discussion topics for dual-stance users' tweets. Lighter shade bars represent the expected number of tweets, and darker bars represent the actual observation. (a) For all dual-stance users, (b) for anti-leaning users with $p_i >= 0.9$, (c) for balanced users with $p_i >= 0.9$, (d) for pro-leaning users with $p_i >= 0.9$.}
\end{figure*}

In order to develop a comprehensive understanding of the content posted by the dual-stance users and address Research Question~2, we classify the tweets into different discussion topics using the composite topic modelling method similar to \cite{Zaidi2023}, and present the contrast between expected and observed tweets' count for each topic. Please see Ref.~\cite{Zaidi2023} for details of the method. 
Figure \ref{topics_all} shows the number of tweets from all dual-stance users classified into 25 most significant anti-vax and pro-vax topics; these topics were identified in \cite{Zaidi2023}, which also provides complete details about how these topics are defined. The light indigo bars represent the expected number of tweets for each anti-vax (pro-vax) topic, calculated by multiplying the total topic tweets by the fraction of anti-vax (pro-vax) tweets from dual-stance users to total anti-vax (pro-vax) tweets in the dataset~\cite{Zaidi2023}. Some topics commonly appeared in both anti- and pro-vax tweets (e.g., `masks and lockdowns', `Pfizer/Moderna vaccines') with opposing perspectives and are shown with indigo coloured labels. Dual-stance users, in general, contributed more tweets than expected for many anti-vax and pro-vax topics, except `jokes (side effects)' and `jokes' as shown in Fig.~\ref{topics_all}. The dual-stance users appear to be highly engaged contributors in the vaccine discourse, who do not send out jokes as much. In contrast, \cite{Zaidi2023} found that pure anti-vax users were far more prominent in spreading anti-vax memes and jokes. 

To gain a more nuanced understanding, we divide the dual-stance cohort into three groups, as described in Section \ref{sec:class}, in terms of their general orientation towards vaccination, i.e., pro-leaning, anti-leaning, or balanced. We only considered the users with 90\% or more probability to be dual-stance, which should lower the impact of misclassified tweets. A majority of the users in this group were pro-leaning at 243,583 (66\%), with 90,313 (25\%) anti-leaning and 32,226 (9\%) balanced. The tolerance value $\epsilon$ is set at $0.05$ (see Section~\ref{sec:class}).

The topics for anti-leaning users are in Fig.~\ref{topics_Anti}. This group was highly active in all anti-vax topics, except `jokes (side effects)', `jokes', and `Trump'. This group contributed less than expected to the top pro-vax topic of `vaccine development'. However, they showed higher activities in `masks and lockdown', `viral infections', `Fauci', `big pharma', `Bill Gates', and `supporting vaccine mandate'. This group seemed to be not in favour of vaccines generally, but also shared some vaccine updates and seemed to be aware of the positive impact of COVID-19 responses, such as masks, lockdowns, and vaccines. They were highly involved in all discussions related to COVID-19, including pandemic control measures where they opposed and supported masks and lockdowns, posted tweets with updates about Bill Gates' efforts in vaccine development as well as criticized him, put forward arguments against vaccine mandates but also advocated for them.

The results for topic classification for the balanced group appear in Fig.~\ref{topics_bal}. Since we are only analysing users with $90\%$ or more likelihood to be dual-stance, $43\%$ users in the balanced class were excluded as they only posted one tweet each with the opposite stances. The selected balanced users each tweeted between 7 and 514 tweets in total, equal or slightly more anti-vax than pro-vax tweets (see precision for anti- and pro-vax classes in Section~\ref{sec:prep}).


The balanced group posted a higher than expected number of tweets about a number of topics, such as, `masks and lockdowns', `Fauci', and `Bill Gates' from both anti- and pro-vax perspectives, `debunking anti-vax', `pro-vax (general)', `Trump', `viral infections', `elections', `no need for vaccine', etc. Unlike the rest of the dual-stance cohort, this group took a greater part in the topics of `jokes (side effects)' and `jokes'. They appear to be aware of current affairs and occasionally share posts on such topics. They seem to be supporting vaccines but also sharing a lot of jokes and memes about vaccines and their side effects, probably forwarding and responding to online content they found amusing without adhering to a particular opinion or taking it seriously. Moreover, some of these users could simply be undecided about the vaccine. 

Finally, the pro-leaning group presents a mirror image of the anti-leaning group. They showed lower than expected activity in most of the anti-vax topics, except for `Pfizer/Moderna', `Trump', and `AstraZeneca (AZ)' and higher than expected contributions in most of the pro-vax topics. The AstraZeneca vaccine made news headlines in March 2021 when several European countries suspended its use following blood clot reports among vaccinated individuals~\cite{AZ-suspend}. In our dataset, people who were supportive of vaccinations also expressed concerns about the AZ vaccine. This pro-leaning group appears to have been looking forward to COVID-19 vaccines, but at the same time are concerned about their safety.

\subsection{Genuine Concerns vs.\ Falsehoods}

\begin{figure}[tbh!]
\begin{center}
\includegraphics[width= 0.46\textwidth,height=6.5cm, clip = true]{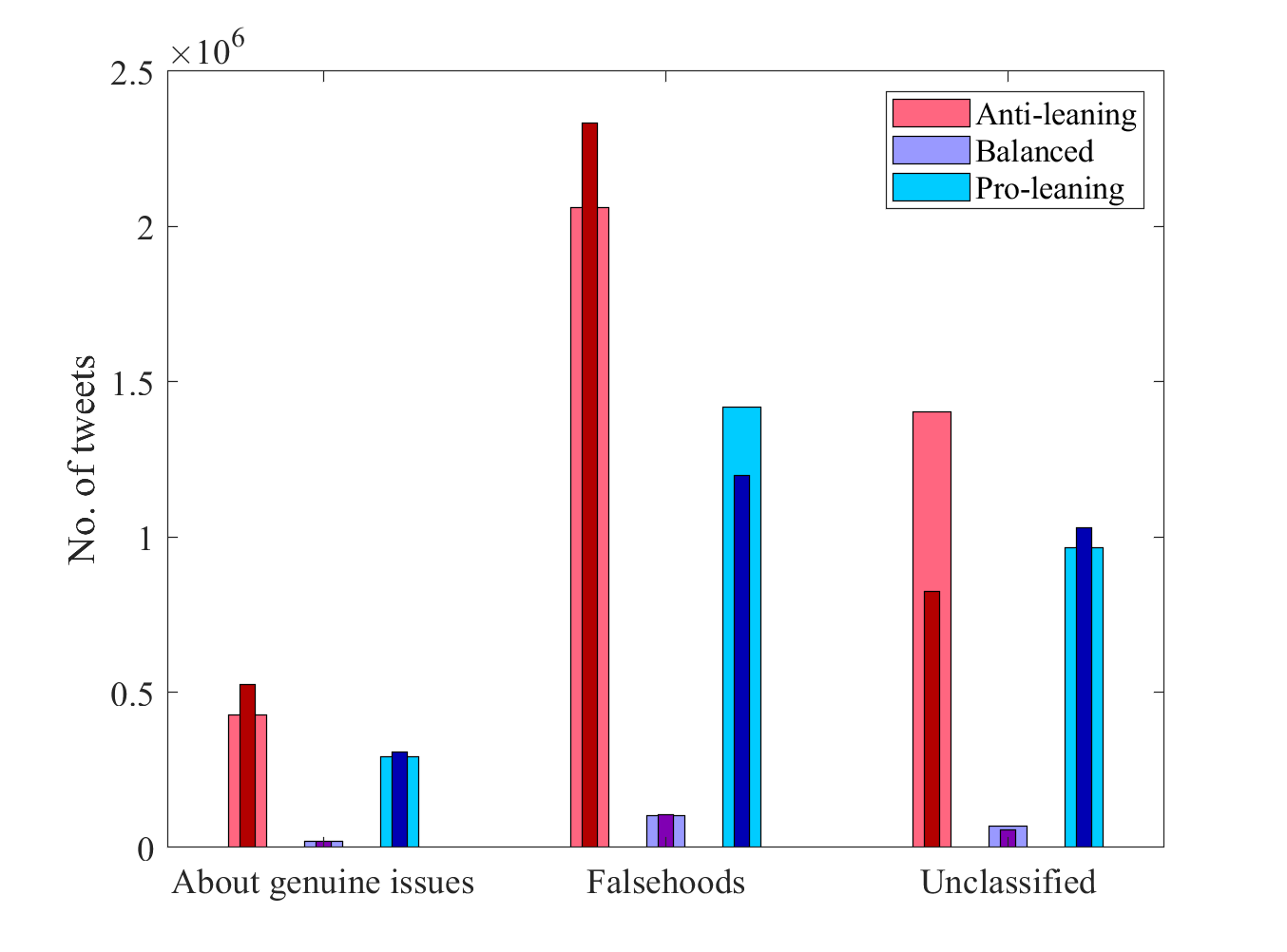}
\caption{Genuine issues versus falsehoods classification of anti-vax tweets.}
\label{dual-legit}
\end{center}
\end{figure}

Figure~\ref{dual-legit} gives another perspective, where we classified tweets into genuine issues or falsehoods, see \cite{Zaidi2023} for details. While genuine issues refer to concerns about unknown long-term side effects of new vaccines, etc., falsehoods are mostly fictitious issues such as claiming that COVID-19 is a hoax for more control and profit (see \cite{Zaidi2023} for details).    
There were some tweets which remained unclassified because of lack of relevant keywords, or they discussed neither genuine issues nor falsehoods. 
Here, we show the expected number of tweets for each class with wider light-coloured bars and overlay it with the observed number of tweets with narrower and darker bars. 
Figure~\ref{dual-legit} shows that the anti-leaning group has more than the expected number of tweets touching on genuine issues, but even more tweets with falsehoods. Following from Fig.~ \ref{topics_Anti}, where anti-leaning users show more than expected contribution in the most popular anti-vax topics of `vaccine mandate', `side effects', etc., and less than expected contribution in `jokes'; it seems that these people are typically not hardcore conspiracy-believing anti-vaxxers, but rather anxious and highly concerned about COVID-19 vaccines. Governments and policymakers should therefore place effort into having a genuine dialogue on these issues.

On the other hand, the pro-leaning group has a less-than-expected amount of tweets involving misinformation, but slightly more tweets discussing genuine issues. The balanced group's contributions did not deviate from expectations. The pro-leaning and anti-leaning groups show contrasting behaviours. However, the pro-leaning group's engagement with misinformation, no matter how limited, shows the prevalence and impact of misinformation and points to the importance of developing solutions and counter-strategies.  

\section{Temporal Dynamics and Causality} \label{sec:temporal-dynamics}

\begin{figure}[tbh!]
\begin{center}
\includegraphics[width= 0.46\textwidth,height=7cm, clip = true]{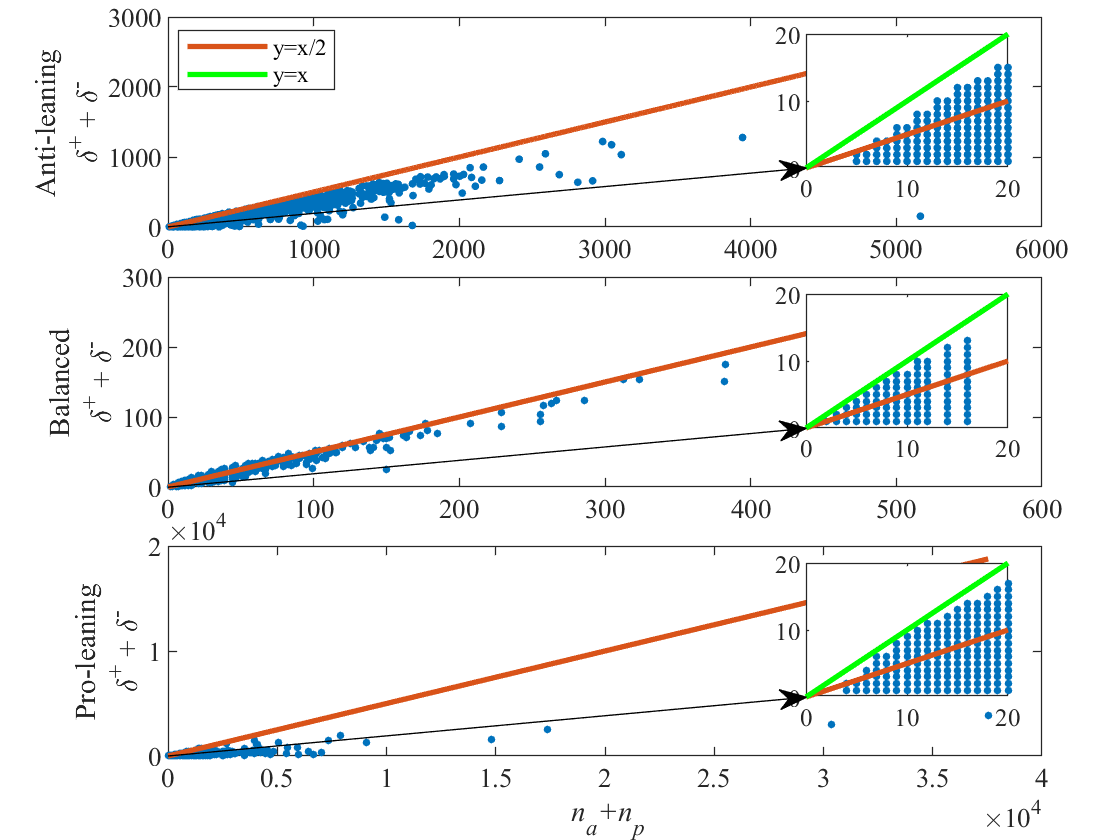}
\caption{Total stance changes versus total anti-vax and pro-vax tweets for anti-leaning, balanced, and pro-leaning user groups. Insets are zoomed-in plots for respective groups.}
\label{fig:change-tweet}
\end{center}
\end{figure}

Many, though not all, active dual-stance users who posted many tweets tended to change their stances many times. In this section, we address Research Question~3 and present our observations regarding the following questions: 1- how frequently, or infrequently, did users change their stances, and 2- is there some correlation between COVID-19 related events and stance change?

We define stance change as a user posting an anti-vax tweet following a pro-vax tweet, or vice versa. Define $\delta^+_i(n)$ as the number of stance changes towards pro-vax and $\delta^-_i(n)$ as the number of stance changes towards anti-vax for a user $i$ on day $n$ during our study period. That is, a pro-vax tweet from a user $i$ on any day at or before $n$ subsequently followed by an anti-vax tweet on day $n$ adds 1 to $\delta^-_i(n)$. Similarly, $\delta^+_i(n)$ counts the pro-vax tweets from a user $i$ on day $n$ which came right after a past anti-vax tweet. We ignored the neutral stance for this analysis. We also define,
\begin{eqnarray}
\delta^+_i = \sum_n \delta^+_i(n)~~\mbox{and}~~\delta^-_i = \sum_n \delta^-_i(n),
\end{eqnarray}
where $\delta^+_i$ and $\delta^-_i$ adds up all stance changes towards and away from pro-vax for a user $i$ over the whole study period. Note that, by construction, 
 $|\delta^+_i - \delta^-_i| \in \{0, 1\}$. 

\begin{figure}
\begin{center}
 \includegraphics[trim={250 100 0 150},width= 0.45\textwidth,height=6.5cm]{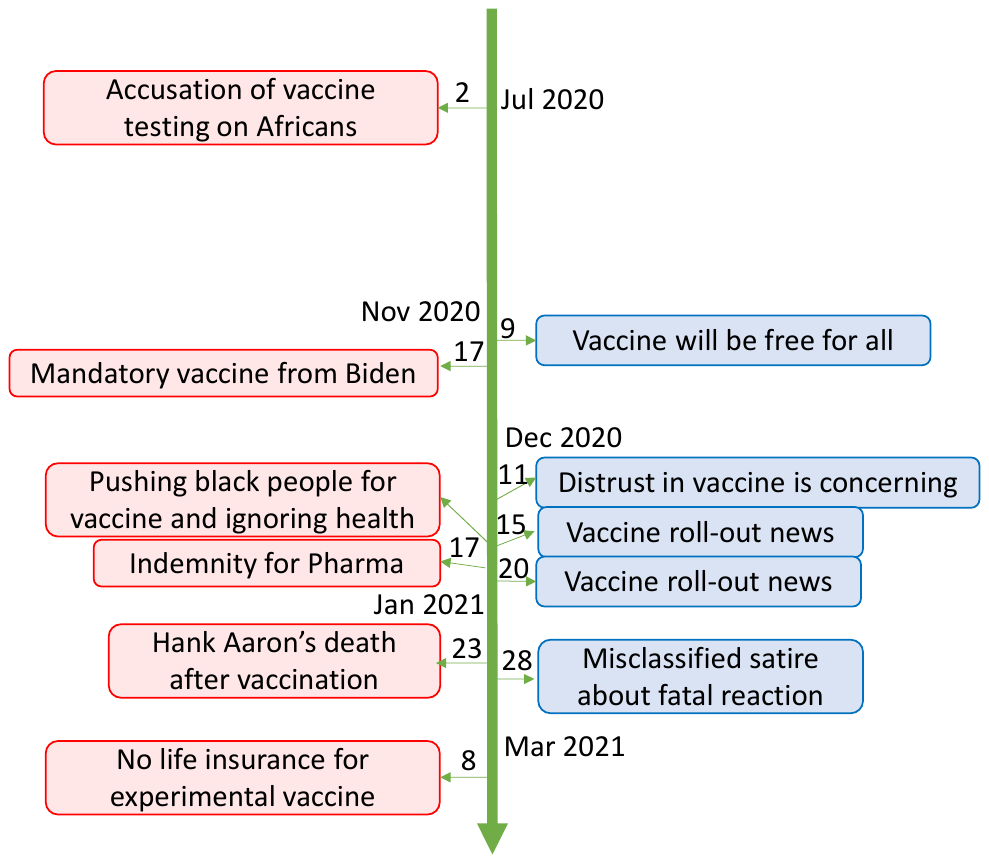}
\caption{An example of a user's timeline who is oscillating between anti-vax and pro-vax stances. Actual tweets are summarised for better visualisation and to avoid user identification.}
\label{fig:oscillating}
\end{center}
\end{figure}

Figure~\ref{fig:change-tweet} shows the number of stance changes for a total number of anti-vax and pro-vax tweets for each individual. The number of stance changes should always be less than the total tweets. However, from the figure, it seems that fewer than half the tweets from a given user involve a change of stance, because most of the scatter points are under the red line (which indicates a user that changes stance in half of their tweets). For users with a smaller number of tweets, shown in the inset charts, stance changes are close to the total number of tweets shown with the green line in the inset chart. This indicates an actual oscillating behaviour between anti- and pro-vax stances. Anti-leaning users appear to have fewer oscillations, but it is because the users with only a few more anti-vax tweets than pro-vax tweets are assigned to the balanced or pro-leaning groups due to low precision for detecting the anti-vax stance. An example of an oscillating user is shown in Fig.~\ref{fig:oscillating}. This user is classified as balanced, although there are more anti-vax tweets than pro-vax. Most of the oscillatory behaviour happened after Nov. 9, 2020, when Pfizer announced the interim trial results for their COVID-19 vaccine \cite{Pfizer-announcement}. We call this post-vaccine period and we will see later that most of the stance changes happened in this period. Moreover, anti-leaning and pro-leaning groups have some users with many tweets but only a few stance changes; however, stance changes in the balanced group stayed closer to the half gradient red line. It reflects the fact that the dataset does not contain users who posted anti-vax tweets for some time, then switched to pro-vax stance and stayed pro-vax for some time. The oscillating behaviour appears to be a characteristic of the balanced group.

In addition, we define $\delta^+(n)$ and $\delta^-(n)$, which count stance changes over all dual-stance users for each day, i.e., 
\begin{eqnarray}
\delta^+(n) = \sum_i \delta^+_i(n)~~\mbox{and}~~\delta^-(n) = \sum_i \delta^-_i(n).
\end{eqnarray}

\begin{figure}[tbh!]
\begin{center}
\includegraphics[width= 0.5\textwidth,height=6.5cm, clip = true]{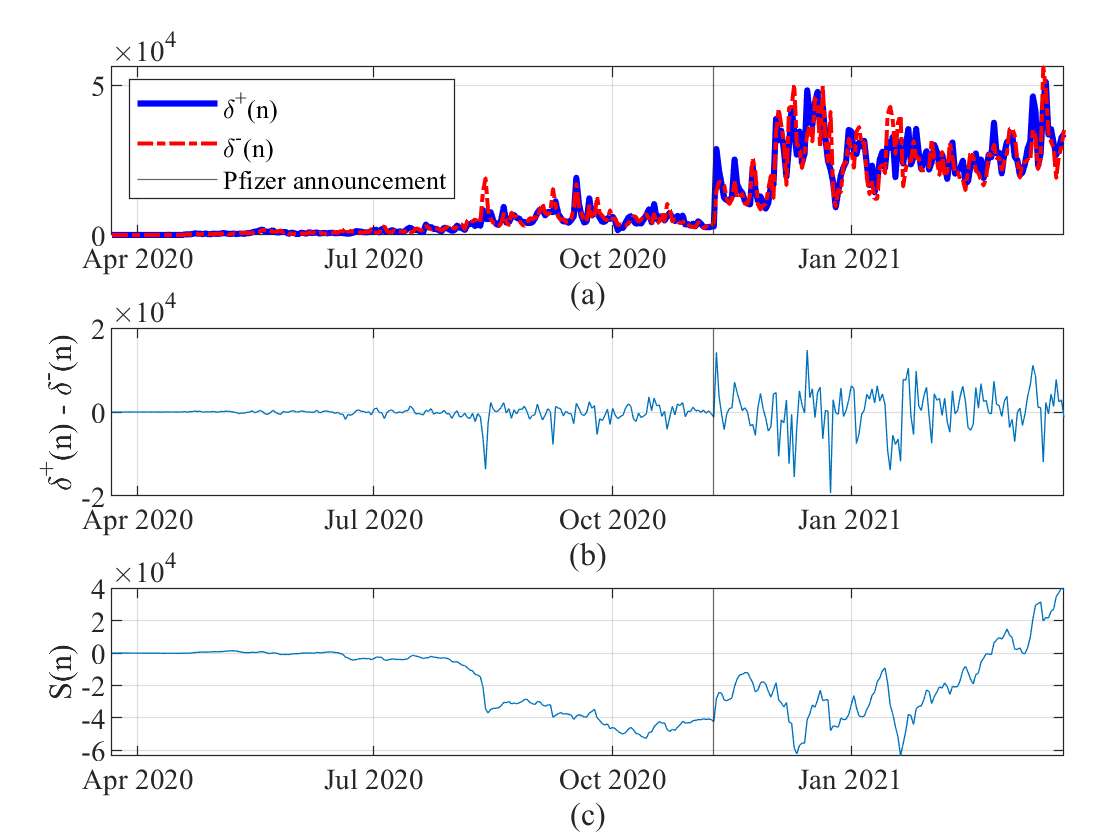}
\caption{(a) Number of changes in stance towards anti-vax and pro-vax, (b) their relative difference, and (c) $S(n) = \sum_1^n(\delta^+(n) - \delta^-(n))$ for each day. The black solid line marks the Pfizer announcement on Nov 9, 2020.}
\label{time-change}
\end{center}
\end{figure}

Figure~\ref{time-change}a shows $\delta^+(n)$ and $\delta^-(n)$ for each day $n$ in our study period. The solid black line marks the announcement of Pfizer's preliminary trial results on Nov. 9, 2020 \cite{Pfizer-announcement}. Interestingly, both trajectories largely overlap and appear to be correlated. Later, we will explore this correlation in detail, but now, let us focus on some observations from these plots. The Pfizer announcement clearly marks the start of a higher activity period, which continued until the end of our study period. The bottom curve Fig.~\ref{time-change}b shows the difference of $\delta^+(n) - \delta^-(n)$ for each day, where we can see that right after the Pfizer announcement, there is a peak towards the pro-vax stance followed by many dips and peaks. Although, Fig.~\ref{time-change}b shows large dips, i.e., more stance changes towards anti-vax, but the cumulative effect of stance changes moved towards the pro-vax side towards the end of the study, as shown by Fig.~\ref{time-change}c, where $S(n) = \sum_1^n(\delta^+(n) - \delta^-(n))$ is plotted versus $n$. 

The most prominent dip in the pre-vaccine period (before Pfizer's announcement) in Fig.~\ref{time-change}b came on Aug 13, 2020, just after the public announcement about Russia's Sputnik vaccine phase 2 trial results~\cite{Sputnik}, and the second dip came on Sep 8, 2020 when AstraZeneca's phase 3 trials were put on hold after one volunteer developed an unknown reaction~\cite{AZ-phase3}. Dual-stance tweeting on COVID-19 vaccination appears to have been triggered by the announcements of successful trial results of the COVID-19 vaccines. Multiple dips in Fig.~\ref{time-change}b, during December 2021, coincide with the Pfizer-BioNTech and Moderna vaccines gaining various approvals in the United Kingdom (UK) and from the USA's Food and Drug Administration (FDA)~\cite{Pfizer-approved-UK,Pfizer-eua,Pfizer-recommend-FDA,Moderna-recommend-FDA,Moderna-eua}, and the launch of UK's public vaccination program on Dec.~8, 2020.

\subsection{Convergent Cross Mapping (CCM)}

The almost synchronous peaks and valleys of the time series plots in Fig.~\ref{time-change}a suggest a strong correlation between $\delta^+(n)$ and $\delta^-(n)$. Indeed, the correlation coefficient was $0.96$. This prompted a further exploration of potential causal relationships between the daily changes to pro- or anti-vax stances in the Twittersphere.

We used Convergent Cross Mapping (CCM) \cite{Sugihara2012}, a non-parametric method based on the theory of nonlinear dynamical systems, to infer causality between two variables $x$ and $y$ based on their time-series data. CCM follows Takens’ theorem, which states that that if indeed $x$ influences $y$, with $x$ and $y$ being state variables of a dynamical system, then past values of $x$ can be recovered from $y$~\cite{Ye2015}. Ref.~\cite{Sugihara2012} developed the technique of ``cross mapping”, where a time delay embedding from the time series of $y$ is used to estimate the values of $x$ and the causal effect of $x$ on $y$ is determined by how well $y$ cross maps $x$, or how well $x$ is forecasted by $y$ \cite{Ye2015}. The forecasting skill or the capacity of the time series data of a variable to predict the other is used to infer causality. The variable with a higher forecasting skill can be interpreted as being driven or caused by the other variable in a non-linear dynamical system. CCM requires the time series to have statistical stationarity. Since the publication of \cite{Sugihara2012}, CCM has been widely used in life sciences, social networking, and environmental research. 

First, we checked for statistical stationarity of $\delta^+(n)$ and $\delta^-(n)$ by using the Augmented Dickey-Fuller (ADF) test and Kwiatkowski–Phillips–Schmidt–Shin (KPSS) test available as Matlab functions. Both $\delta^+(n)$ and $\delta^-(n)$ were found stationary by the ADF test, but not by the KPSS test. For such cases, the standard approach is to take a first order difference of the time series, which should be stationary. Instead of $\delta^+(n)$ and $\delta^-(n)$, we applied CCM\footnote{\url{https://skccm.readthedocs.io/en/latest/quick-example.html}} algorithm over $\delta^+(n+1) - \delta^+(n)$ and $\delta^-(n+1) - \delta^-(n)$ and calculated their forecast skill or cross-map (xmap) skill to predict each other for different library lengths, ranging from a small value to the full length of the time series. The value for lag is selected using the mutual information of the time series, which has a minimum at lag = 3. The mutual information plot is available in the Supplementary Material. 
The embedding dimension is selected as 32 which gives the best prediction skill.

\begin{figure}[tbh!]
\begin{center}
\includegraphics[width= 0.46\textwidth,height=6.5cm, clip = true]{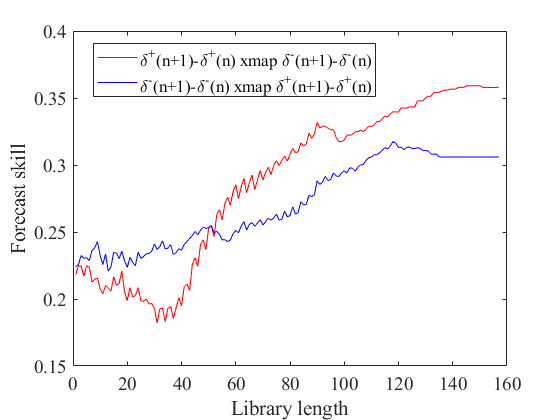}
\caption{Forecast scores calculated from CCM vs. library length, with lag = 3 and embedding dimension = 32.}
\label{ccm}
\end{center}
\end{figure}

The results are given in Fig.~\ref{ccm} for lag = 3 and embedding dimension = 32. A change towards pro-vax stance ($\delta^+(n+1) - \delta^+(n)$) has a higher forecasting skill than a change to anti-vax stance ($\delta^-(n+1) - \delta^-(n)$) for larger library lengths. This indicates that the change to pro-vax stance could be a possible reaction to, or be driven by, the change to anti-vax stance. It is beyond the scope of this study to verify what exactly caused this behaviour but we conjecture the following might have contributed. First, dual-stance users might be posting pro-vax tweets just to respond to, or compensate for, early posting of anti-vax posts, like a balancing act. Second, mainstream media and public health organizations often ran public awareness campaigns in response to conspiracy theories and misinformation about COVID-19 vaccine, which also got traction over Twitter. However, more detailed analysis will be needed to understand if in fact these awareness campaigns have caused the behaviour depicted by CCM in Fig.~\ref{ccm}. 

\section{Influence of Retweets and Replies}\label{sec:RnR}

\begin{figure}{
 \subfloat[]{
    \includegraphics[width= 0.49\textwidth, clip = true]{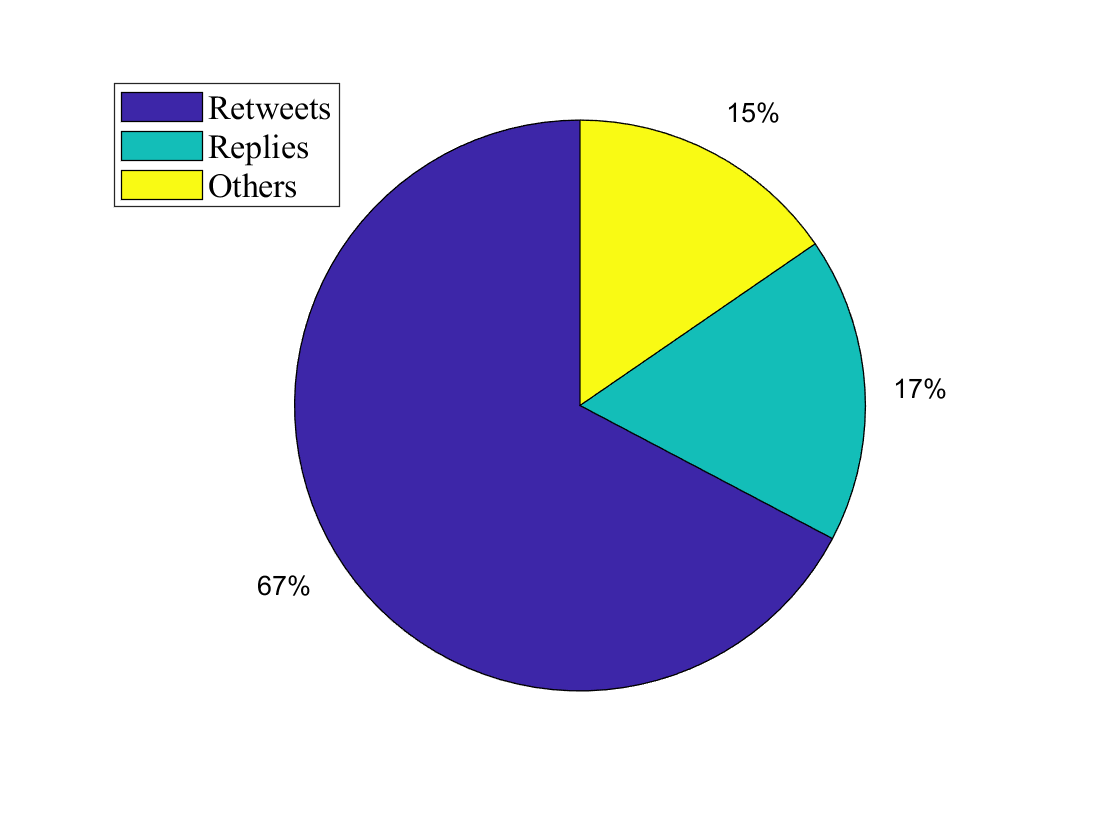}
    \label{fig:change-composition}}\\
 \subfloat[]{
    \includegraphics[width= 0.49\textwidth, clip = true]{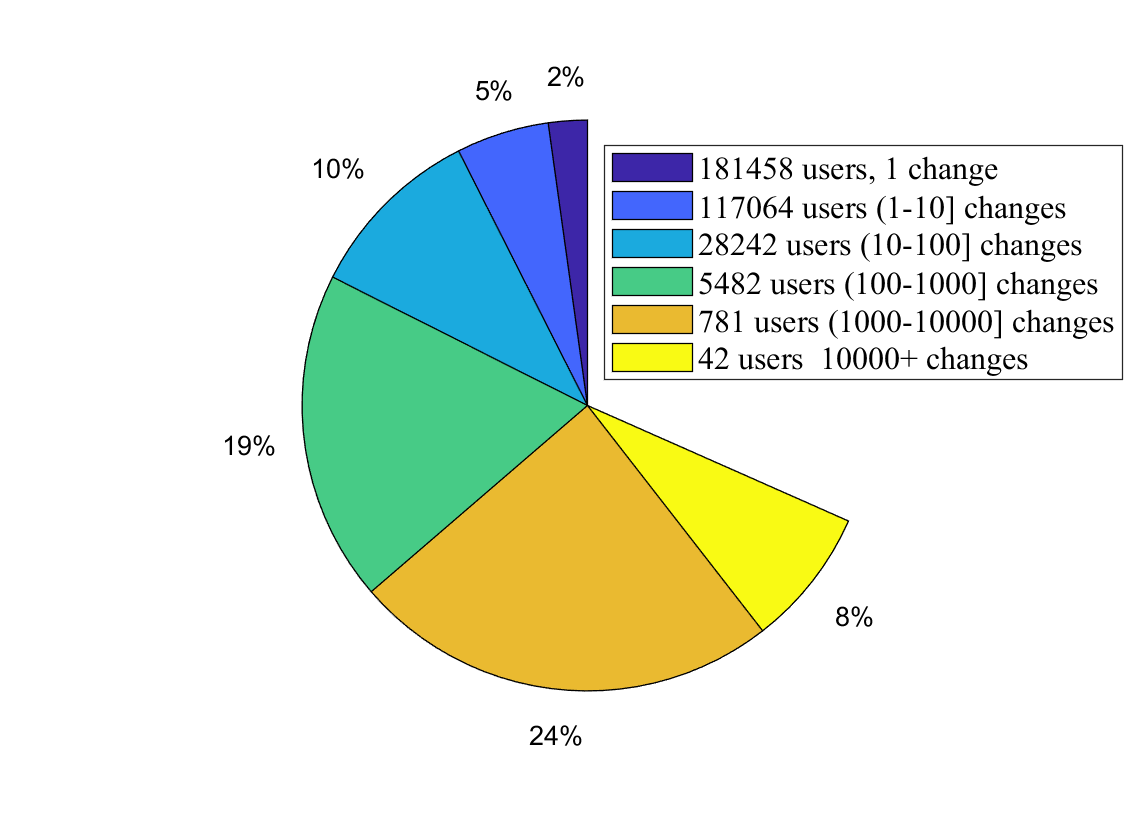}
    \label{fig:change-sources}}
    \caption{(a) The composition of change-tweets: 67\% are retweets and 17\% are replies. (b) Many retweet and reply threads are originated by the same user. Pie chart shows the percentage of such retweets and replies which are originated by users associated with 1, 2-10, 11-100, 101-1000, and 1000+ change-retweets. 6200 users are behind more than 50\% of the stance changes.}}
\end{figure}

In this dataset, there are 8,241,666 anti-vax and pro-vax tweets by dual-stance users, where change of stance is detected, i.e., these tweets were posted after a tweet with opposite stance by the same user.  We denote these tweets as the `change-tweets'. This set contains $67\%$ retweets and $17\%$ reply tweets as shown in the Fig.~\ref{fig:change-composition}. Many such retweets originated from the same user, and many such reply tweets belong to the same thread. We found that the change-tweets included 900,984 and 1,141,124 distinct retweet and reply threads, respectively, initiated by 333,069 unique user IDs. For each of these users, we counted change-tweets in their respective retweet and reply threads. 181,458 users had only 1 change-tweet in their threads, collectively they are 2\% of all of the change-tweets. We made 5 groups of users, i.e., 
all users associated with 1, (1 10], (10 100], (100 1000] and greater than 1000 change-tweets. In other words, the users were grouped based on how many change-tweets their threads contained. The pie chart in Fig.~\ref{fig:change-sources} shows the percentage of change-tweets associated to each user group. Note that we could not hydrate all retweet/in-reply-to IDs, and the Fig.~\ref{fig:change-sources} shows $68\%$ change-tweets, although as stated above, $67\%$ of these were retweets and $17\%$ were replies. 

The most interesting observation from Fig.\ \ref{fig:change-sources} is that a relatively small proportion of users are associated with numerous change-tweets, i.e., their tweets or threads have driven numerous stance changes. More precisely, 6,200 users are linked with more than $50\%$ of the stance changes detected in our dataset, and just 42 users, out of 333,069 unique user IDs, contributed $8\%$ of the total stance changes. 
Supplementary Material contains further details about these 42 users, and also a detailed topic analysis of the retweets where stance changes occurred.

\begin{figure}[tbh!]
\begin{center}
\includegraphics[trim={0 0 300 0},width= 0.4\textwidth,height=5cm, clip = true]{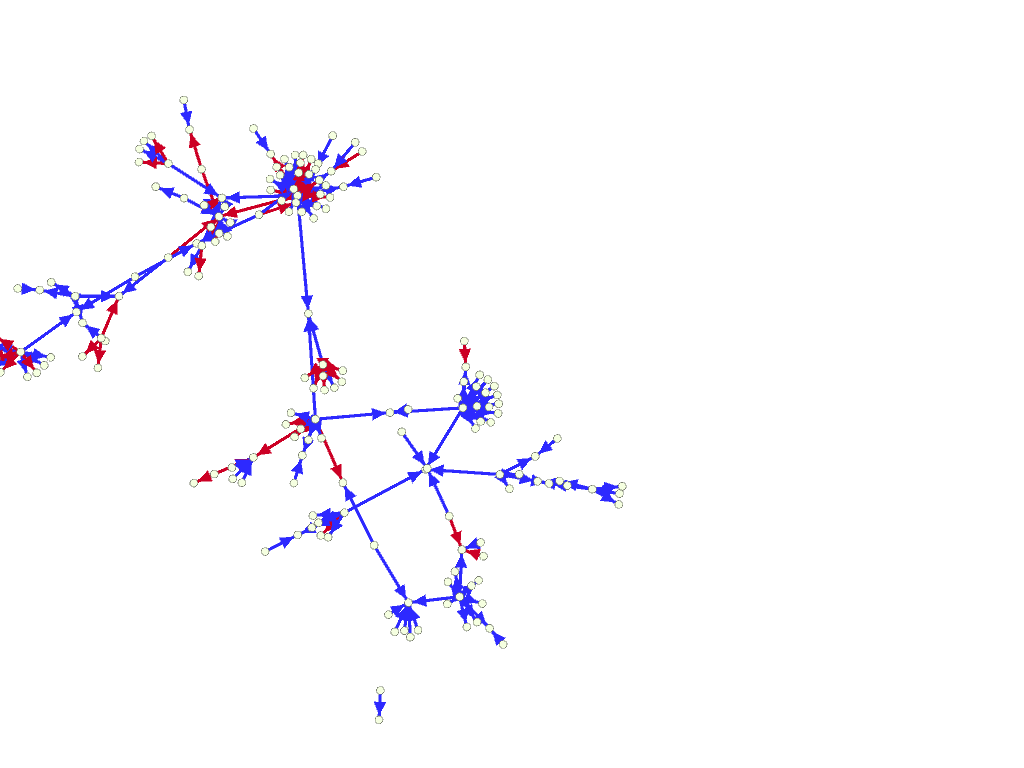}
\caption{A connected component from `Reply' graph for dual-stance users changing their stances on Nov 9, 2020. Directed link are from the originator of the thread to  the replying user. The red colour represents that both original tweet and reply tweet have opposite stances, and a blue colour represents the same stances.}
\label{fig:reply-graph}
\end{center}
\end{figure}

In COVID-19 vaccine related tweets, reply threads are mostly active discussions between the two opposite camps. Figure~\ref{fig:reply-graph} shows a connected component from the reply graph from the change-tweets posted on Nov 9, 2020, the first day of the post-vaccine period, when preliminary results of the Pfizer vaccine were announced \cite{Pfizer-announcement}. Edges represent replies in Fig.~\ref{fig:reply-graph}, and they are directed links from the users who posted the original tweet (source nodes) to the replying users (target nodes). Figure \ref{fig:reply-graph} has 186 edges and there are 53 root nodes. We consider a signed graph, whereby an edge can have a +1 (blue) or -1 (red) weighting. For the reply network, the weight of an edge between two nodes is +1 (-1) if the tweet stances of the two nodes are the same (opposite). In the reply graph in Fig.~\ref{fig:reply-graph}  with red edges, we can see some debates and exchange of opposing views. Hundreds of such reply threads are found in the change-tweets, and more details are presented in the Supplementary Material. 

Replies and retweets have different social implications. Retweets are used primarily as a form of agreement, whereas replies may be more prevalent for debates and discussions with differing views. We also want to raise a point here; even if there are some clusters or echo chambers of like-minded users, the presence of such reply networks suggest that these echo chambers have `leaky walls' and people are potentially engaged in genuine discussions across opposing opinion camps. 

\section{Two Years On}


\begin{figure}
\centering
    \subfloat[]{
        \includegraphics[trim={0 0 0 0},width= 0.45\textwidth]{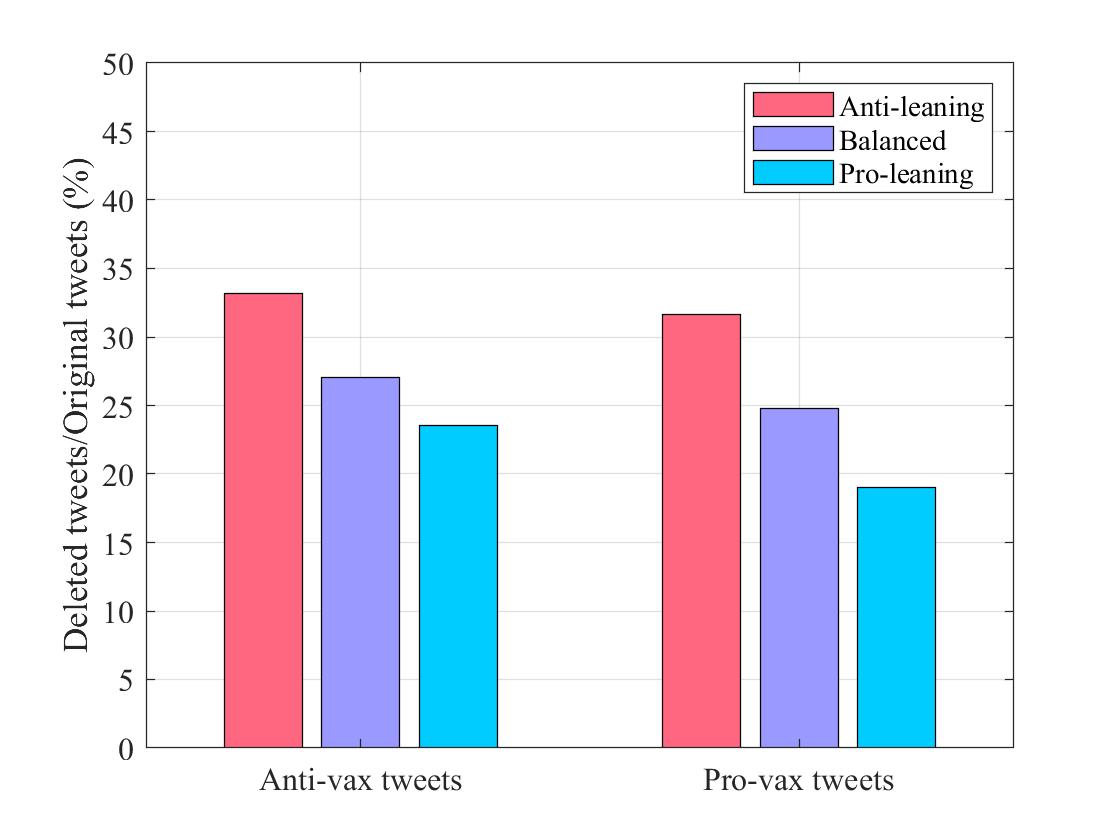}
        \label{fig:tweet-retention}}
        \\
        \subfloat[]{
        \includegraphics[trim={30 120 200 70},width= 0.48\textwidth]{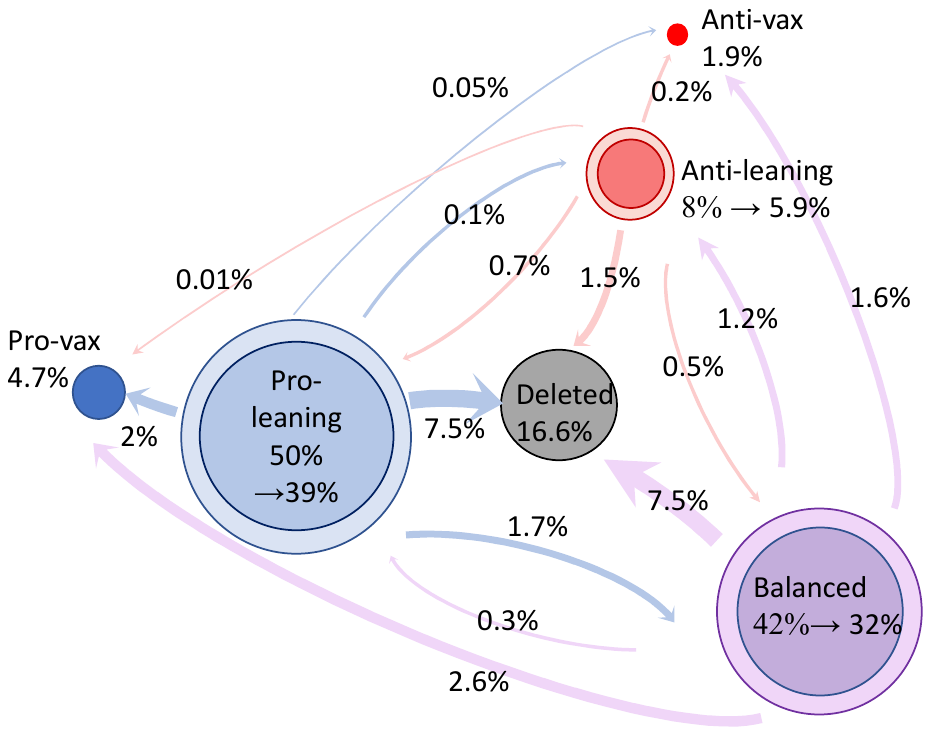}
        \label{fig:stance-rehydration}}
        \caption{In the course of 2 years, early 2021 to early 2023, many of the anti-vax and pro-vax tweets from the dual-stance users went offline. (a) Number of deleted anti-vax (pro-vax) tweets as a percentage of original anti-vax (pro-vax) tweets from the user group. All groups have more deleted anti-vax tweets than pro-vax tweets, this difference is more significant in pro-leaning group. (b) User classification with original data (lighter circles), with rehydrated data (darker circles), and the migration of users to different classes as a result of selective deletion of tweets. Two smaller classes of pure anti-vax and pure pro-vax also emerged. Percentages are calculated with respect to the total number of original dual-stance users.}\vspace{-5mm}
\end{figure}

The dataset analysed in the paper was hydrated in 2021. Twitter is a dynamic platform, and it is quite common that tweets are being deleted or taken offline by the author or by the platform. Authors may delete the post because they have changed their opinions or for other many reasons. The platform usually take tweets or users offline if they are found to be in breach of the code of conduct on Twitter. An interesting statistic is the number of tweets deleted since our hydration. A user's choice to delete tweets of one stance while retaining tweets of the opposite stance may reflect the fact that one stance has prevailed over the other, as time has gone by. To investigate this, we hydrated the same set of tweets again and recorded which tweets had been deleted or taken offline

For the overall dual-stance cohort, we observed that $28\%$ anti-vax tweets have been taken offline since we first hydrated the dataset, whereas $20\%$ pro-vax tweets were deleted. If we look at comparable statistics in the anti-leaning, balanced, and pro-leaning groups, it is found that the anti-leaning group has the highest percentage of deleted tweets (both anti- and pro-vax tweets) as shown in Fig.~\ref{fig:tweet-retention}. However, the difference between deleted anti-vax and pro-vax number of tweets is greatest in the pro-leaning group.

Figure~\ref{fig:stance-rehydration} shows the change in user classification in relation to the retained tweets. Lighter circles show the original user classes in terms of fraction of dual-stance users assigned to it, and darker circles show the classification using the retained tweets. All classes have shrunk in size, but the majority of the users retained their labels. Interestingly, among those who retained their original label, $37\%$ of anti-leaning, and $47\%$ each of balanced and pro-leaning users also deleted some of their tweets, but the deletion did not change the relative composition of dual-stance tweets.

Out of the users who moved to a different class, a majority have deleted all anti-vax and pro-vax tweets, as shown by the grey circle in the middle. We have not checked, but some of these users in the grey circle might also be offline or banned from Twitter. The second largest migrating pro-leaning group became pure pro-vax, by deleting all of their anti-vax tweets. The emerging pure pro-vax class got an even greater movement from the balanced group and a minuscule portion from the anti-leaning group. Similarly, a tiny class of pure anti-vax also emerged, mostly out of the balanced group and with a small contribution from anti-leaning and, even smaller from pro-leaning groups. 

Users moving away from the balanced group by selectively deleting their tweets are almost evenly split between pro-vax/pro-leaning and anti-vax/anti-leaning, as shown in Fig.~\ref{fig:stance-rehydration}. Interestingly, the majority of the users with selective deletion from anti-leaning moved to pro-leaning, while some users also moved to the balanced group. With selective deletion, majority of pro-leaning users became pro-vax and balanced while a tiny fraction of them became anti-leaning or anti-vax.  


If the oscillation between the stances is an indication of the uncertainty of the time, then a higher deletion of anti-vax posts may indicate an act of rectification from some users. An overall trend in users who have selectively deleted their tweets is to favour and maintain pro-vax stance tweets. The balanced group, however, remains a puzzling phenomenon with an almost even split between favouring anti-vax and pro-vax stances.    

\section{Discussion and Conclusion}\label{sec:conclusion}

The expression of an opinion and its interpretation are noisy processes. To make things more difficult, in the Twittersphere, opinions are expressed through a message with a maximum of 280 characters. In the present study, the stances of tweets were automatically detected by state-of-the-art transformer models, which are still outperformed by humans in this task. Therefore, one would expect some noise in the stance detection. Nonetheless, even when the limitations of our methods are considered, we found a significant cohort of Twitter users who provided arguments both for and against COVID-19 vaccination. One might expect that there would be a 'fuzzy group' of this type -- those who cannot be clearly classified as anti-vaxxers or pro-vaxxers. However, the size of this group -- 22\% of the users -- along with their significant contributions to the overall number of tweets, is surprising and perhaps raises a question as to how polarised the vaccine debate truly is.

In this paper, we attempted to gain an understanding and further insight into this dual-stance cohort. Our content analysis shows that this cohort can be divided into 3 groups: 1 anti-leaning, 2- balanced, and 3- pro-leaning. The anti-leaning group is highly active in sharing concerns about COVID-19 vaccines, including many falsehoods and misinformation. They shared some vaccine updates and positive news, but they appear to not be in favour of COVID-19 vaccines. The balanced group is significant in size, and seems to be on the fence as far as the vaccine is concerned. They are undecided, sharing news of vaccine developments while worried about safety and mandating of COVID-19 vaccines. The pro-leaning group is most numerous and appears to be favouring COVID-19 vaccines, though this group does present some concerns about them. A large number of tweets that indicated a change of stance are retweets and replies. Also, both anti-vax tweets and pro-vax tweets are found to be part of many reply threads, indicating that people engaged with others with opposing views and changed their opinions.

The temporal analysis of change of stance suggests a strong correlation between change to anti-vax stance and change to pro-vax stance. Significant activity in terms of volume of stance changes occurred after the public release of clinical trial results of various COVID-19 vaccines. The change towards a pro-vax stance also seemed to be a reaction to the change towards an anti-vax stance, as suggested by our analysis using the Convergent Cross Mapping method. There are days when overwhelming volume of changes of stance took place towards anti-vax, for example, when AstraZeneca trials were suspended due to a participant developing an unknown reaction. However, towards the end of our study period, the cumulative effect was positive, and stance changes were mostly towards the pro-vax stance.  

In this study, we explored a surprising and perplexing phenomenon, namely the major presence of dual-stance users. Our analyses provided insight into their behaviour pattern and painted a comprehensible and plausible picture. The binary division between anti-vax and pro-vax classes was perhaps too strict for many people, especially during the uncertain and unprecedented time of the COVID-19 pandemic. In fact, the idea of dual-stance seems almost similar to the concept of political ambivalence \cite{Webb2018}, where not only are there contrasting and ambivalent views about political left or right within a population, any one individual could also hold contrasting opinions within themselves and partially subscribe to one view or partially reject the other, or just remain undecided. There have been many efforts recently to understand the major drivers and mindsets of this group and their role in democracies. Within our context of COVID-19 vaccines, the positive point of this dual-stance cohort, which exists at the boundary of polarising camps, is that it can act as a bridge and bring opposing communities together. The need, however, is to create opportunities for genuine dialogue where legitimate concerns can be acknowledged and addressed and trust can be established between people and public health organizations.

\section*{Acknowledgement}

We would like to thank Dr. Marc Cheong for his insightful comments about this work.

\bibliography{reference,References2}

\begin{thebibliography}{10}

\bibitem{Zaidi2023}
Z.~Zaidi, M.~Ye, F.~J. Samon, A.~Jama, B.~Gopalakrishnan, C.~Gu, S.~Karunasekera, J.~Evans, and Y.~Kashima, ``Topics in anti-vax and pro-vax discourse: A yearlong synoptic study of covid-19 vaccine tweets,'' {\em Journal of Medical Internet Research (in press)}, 2023.

\bibitem{Rathje2022}
S.~Rathje, J.~K. He, J.~Roozenbeek, J.~J. Van Bavel, and S.~van der Linden, ``Social media behavior is associated with vaccine hesitancy,'' {\em PNAS Nexus}, vol.~1, p.~pgac207, Sept. 2022.

\bibitem{DeNicola2023}
G.~De~Nicola, V.~H. Tuekam~Mambou, and G.~Kauermann, ``{COVID-19 and social media: Beyond polarization},'' {\em PNAS Nexus}, vol.~2, p.~pgad246, 08 2023.

\bibitem{Monsted2022}
B.~Mønsted and S.~Lehmann, ``Characterizing polarization in online vaccine discourse—a large-scale study,'' {\em PLOS ONE}, vol.~17, pp.~1--19, 02 2022.

\bibitem{Jiang2021}
X.~Jiang, M.-H. Su, J.~Hwang, R.~Lian, M.~Brauer, S.~Kim, and D.~Shah, ``Polarization over vaccination: Ideological differences in twitter expression about covid-19 vaccine favorability and specific hesitancy concerns,'' {\em Social Media + Society}, vol.~7, no.~3, p.~20563051211048413, 2021.

\bibitem{Gori2021}
D.~Gori, F.~Durazzi, M.~Montalti, Z.~D. Valerio, C.~Reno, M.~P. Fantini, and D.~Remondini, ``Mis-tweeting communication: a vaccine hesitancy analysis among twitter users in italy,'' {\em Acta Biomed.}, vol.~92, p.~e2021416, Oct 2021.

\bibitem{Crupi2022}
G.~Crupi, Y.~Mejova, M.~Tizzani, D.~Paolotti, and A.~Panisson, ``Echoes through time: Evolution of the italian covid-19 vaccination debate,'' {\em Proceedings of the International AAAI Conference on Web and Social Media}, vol.~16, pp.~102--113, May 2022.

\bibitem{Giovanni2022}
M.~D. Giovanni, F.~Pierri, C.~Torres-Lugo, and M.~Brambilla, ``Vaccineu: Covid-19 vaccine conversations on twitter in french, german and italian,'' {\em Proceedings of the International AAAI Conference on Web and Social Media}, vol.~16, pp.~1236--1244, May 2022.

\bibitem{Poddar2022}
S.~Poddar, M.~Mondal, J.~Misra, N.~Ganguly, and S.~Ghosh, ``Winds of change: Impact of covid-19 on vaccine-related opinions of twitter users,'' {\em Proceedings of the International AAAI Conference on Web and Social Media}, vol.~16, pp.~782--793, May 2022.

\bibitem{Weinzierl2022}
M.~A. Weinzierl and S.~M. Harabagiu, ``From hesitancy framings to vaccine hesitancy profiles: A journey of stance, ontological commitments and moral foundations,'' {\em Proceedings of the International AAAI Conference on Web and Social Media}, vol.~16, pp.~1087--1097, May 2022.

\bibitem{Ebeling2022}
R.~Ebeling, C.~A.~C. Saenz, J.~C. Nobre, and K.~Becker, ``Analysis of the influence of political polarization in the vaccination stance: The brazilian covid-19 scenario,'' {\em Proceedings of the International AAAI Conference on Web and Social Media}, vol.~16, pp.~159--170, May 2022.

\bibitem{Kleitman2023}
S.~Kleitman, D.~J.~F. DJ, M.~K.~H. Law, M.~D. Blanchard, R.~Campbell, M.~A. Tait, J.~Schulz, J.~L. amd L.~Stankov, and M.~T. King, ``The psychology of covid-19 booster hesitancy, acceptance and resistance in australia,'' {\em Vaccines (Basel)}, vol.~11, April 2023.

\bibitem{Lamsal2021}
R.~Lamsal, ``Design and analysis of a large-scale covid-19 tweets dataset,'' {\em Applied Intelligence}, vol.~51, no.~5, pp.~2790--2804, 2021.

\bibitem{Rao2019}
P.~Rao, ``Transfer learning in nlp for tweet stance classification.'' \url{https://towardsdatascience.com/transfer-learning-in-nlp-for-tweet-stance-classification-8ab014da8dde}, 2019.
\newblock Accessed: 05-07-2021.

\bibitem{AZ-suspend}
ABC-News, ``{AstraZeneca COVID-19 shots stopped by several European nations amid blood clot reports, but EU regulator says no evidence of link}.'' \url{https://www.abc.net.au/news/2021-03-12/denmark-norway-iceland-suspend-astrazeneca-covid-shots-vaccine/13240984}, 2021.
\newblock accessed on 04-04-2022.

\bibitem{Pfizer-announcement}
J.~Gallagher, ``{Covid vaccine: First `milestone' vaccine offers 90\% protection}.'' \url{https://www.bbc.com/news/health-54873105}, 2020.
\newblock accessed on 04-04-2022.

\bibitem{Sputnik}
``Clinical trials.'' \url{https://sputnikvaccine.com/about-vaccine/clinical-trials/}, 2021.
\newblock accessed on 04-04-2022.

\bibitem{AZ-phase3}
D.~Beasley, ``{AstraZeneca suspends leading COVID-19 vaccine trials after a participant's illness}.'' \url{https://www.reuters.com/article/us-health-coronavirus-astrazeneca-idUSKBN25Z392}, 2020.
\newblock accessed on 04-04-2022.

\bibitem{Pfizer-approved-UK}
S.~Boseley and J.~Halliday, ``{UK approves Pfizer/BioNTech Covid vaccine for rollout next week This article is more than 1 year}.'' \url{https://www.theguardian.com/society/2020/dec/02/pfizer-biontech-covid-vaccine-wins-licence-for-use-in-the-uk}, 2020.
\newblock accessed on 04-04-2022.

\bibitem{Pfizer-eua}
J.~Howard, ``{FDA issues emergency use authorization for Pfizer/BioNTech Covid-19 vaccine}.'' \url{https://edition.cnn.com/2020/12/11/health/covid-vaccine-fda-eua/index.html}, 2020.
\newblock accessed on 04-04-2022.

\bibitem{Pfizer-recommend-FDA}
``{FDA advisory panel endorses Pfizer/BioNTech Covid-19 vaccine}.'' \url{https://www.statnews.com/2020/12/10/tracking-the-fda-advisory-panel-meeting-on-the-pfizer-biontech-covid-19-vaccine/}, 2020.
\newblock accessed on 04-04-2022.

\bibitem{Moderna-recommend-FDA}
A.~Watts, J.~Christensen, and M.~Fox, ``{FDA plans to authorize Moderna's coronavirus vaccine quickly}.'' \url{https://edition.cnn.com/2020/12/17/health/moderna-covid-vaccine-fda-vrbpac-vote/index.html}, 2020.
\newblock accessed on 04-04-2022.

\bibitem{Moderna-eua}
B.~L. Jr., ``{FDA approves second Covid vaccine for emergency use as it clears Moderna's for U.S. distribution}.'' \url{https://www.cnbc.com/2020/12/18/moderna-covid-vaccine-approved-fda-for-emergency-use.html}, 2020.
\newblock accessed on 04-04-2022.

\bibitem{Sugihara2012}
G.~Sugihara, R.~May, H.~Ye, C.~hao Hsieh, E.~Deyle, M.~Fogarty, and S.~Munch, ``Detecting causality in complex ecosystems,'' {\em Science}, vol.~338, no.~6106, pp.~496--500, 2012.

\bibitem{Ye2015}
H.~Ye, E.~R. Deyle, L.~J. Gilarranz, and G.~Sugihara, ``Distinguishing time-delayed causal interactions using convergent cross mapping,'' {\em Nature Scientific Reports}, vol.~5, p.~14750, Oct. 2015.

\bibitem{Webb2018}
A.~Webb, ``In praise of democratic ambivalence,'' {\em Democratic Theory}, vol.~5, pp.~17--36, Dec. 2018.
\newblock Publisher Copyright: {\textcopyright} 2018 Democratic Theory.

\end{thebibliography}
\bibliographystyle{ieeetr}
\end{document}